\documentclass[sigconf]{acmart}
\AtBeginDocument{%
  }

\copyrightyear{2025}
\acmYear{2025}
\setcopyright{acmlicensed}
\acmConference[MM '25] {Proceedings of the 33rd ACM International Conference on Multimedia}{October 27--31, 2025}{Dublin, Ireland.}
\acmBooktitle{Proceedings of the 33rd ACM International Conference on Multimedia (MM '25), October 27--31, 2025, Dublin, Ireland}
\acmISBN{979-8-4007-2035-2/2025/10}
\acmDOI{10.1145/3746027.3754578}

\usepackage{mathrsfs}
\usepackage{amsmath}
\usepackage{subcaption}
\usepackage{booktabs}
\usepackage{multirow}
\usepackage{makecell}
\usepackage{pifont}
\usepackage{xspace}
\usepackage{hhline}
\usepackage{appendix}
\usepackage{color}
\usepackage{soul}
\usepackage[prologue,table]{xcolor}
\usepackage{dashrule}
\usepackage{balance}

\acmSubmissionID{7206}

\begin{document}

\title{CROP: Integrating Topological and Spatial Structures via Cross-View Prefixes for Molecular LLMs}

\author{Jianting Tang}
\affiliation{%
  \institution{University of Science and Technology of China}
  \institution{State Key Laboratory of Cognitive Intelligence}
  \city{Hefei}
  \state{Anhui}
  \country{China}}
\email{jiantingtang@mail.ustc.edu.cn}

\author{Yubo Wang}
\affiliation{%
  \institution{University of Science and Technology of China}
  \institution{State Key Laboratory of Cognitive Intelligence}
  \city{Hefei}
  \state{Anhui}
  \country{China}}
\email{wyb123@mail.ustc.edu.cn}

\author{Haoyu Cao}
\affiliation{%
  \institution{University of Science and Technology of China}
  \institution{State Key Laboratory of Cognitive Intelligence}
  \city{Hefei}
  \state{Anhui}
  \country{China}}
\email{caohaoyu@mail.ustc.edu.cn}

\author{Linli Xu}
\authornote{Corresponding author.}
\affiliation{%
  \institution{University of Science and Technology of China}
  \institution{State Key Laboratory of Cognitive Intelligence}
  \city{Hefei}
  \state{Anhui}
  \country{China}}
\email{linlixu@ustc.edu.cn}


\newcommand{\modelname}{CROP}
\newcommand{\red}[1]{\textcolor{red}{#1}}
\newcommand{\blue}[1]{\textcolor{blue}{#1}}

\begin{abstract}
Recent advances in molecular science have been propelled significantly by large language models (LLMs).
However, their effectiveness is limited when relying solely on molecular sequences, which fail to capture the complex structures of molecules.
Beyond sequence representation, molecules exhibit two complementary structural views: the first focuses on the \textit{topological} relationships between atoms, as exemplified by the graph view; and the second emphasizes the \textit{spatial} configuration of molecules, as represented by the image view. 
The two types of views provide unique insights into molecular structures.
To leverage these views collaboratively,
we propose the \textbf{CRO}ss-view \textbf{P}refixes (\modelname{}) to enhance LLMs' molecular understanding through efficient multi-view integration.
\modelname{} possesses two advantages: (\textit{i}) efficiency: 
by jointly resampling multiple structural views into fixed-length prefixes, it avoids excessive consumption of the LLM's limited context length and allows easy expansion to more views; (\textit{ii}) effectiveness:
by utilizing the LLM’s self-encoded molecular sequences to guide the resampling process, it boosts the quality of the generated prefixes.
Specifically, our framework features a carefully designed SMILES Guided Resampler for view resampling, and a Structural Embedding Gate for converting the resulting embeddings into LLM's prefixes.
Extensive experiments demonstrate the superiority of \modelname{} in tasks including molecule captioning, IUPAC name prediction and molecule property prediction.

\end{abstract}


\maketitle

%
%
\section{Introduction}
LLMs have exhibited remarkable proficiency across diverse domains~\citep{survey}. In the chemical field, particularly in tasks such as molecule captioning and property prediction~\citep{wu2018moleculenet}, LLMs have emerged as promising tools for streamlining research efforts.
As a molecule’s properties are fundamentally determined by its complex structure~\citep{Fedik2022ExtendingML,Yang2019AnalyzingLM}, providing LLMs with accurate structural representations is essential for enhancing their molecular understanding.
However, current LLMs primarily rely on sequence representations like SMILES~\citep{weininger1988smiles} and SELFIES~\citep{krenn2020self} for molecular tasks, 
which are inadequate for capturing complex molecular structures.

\begin{figure}[t!]
  \centering
  \includegraphics[width=\linewidth]{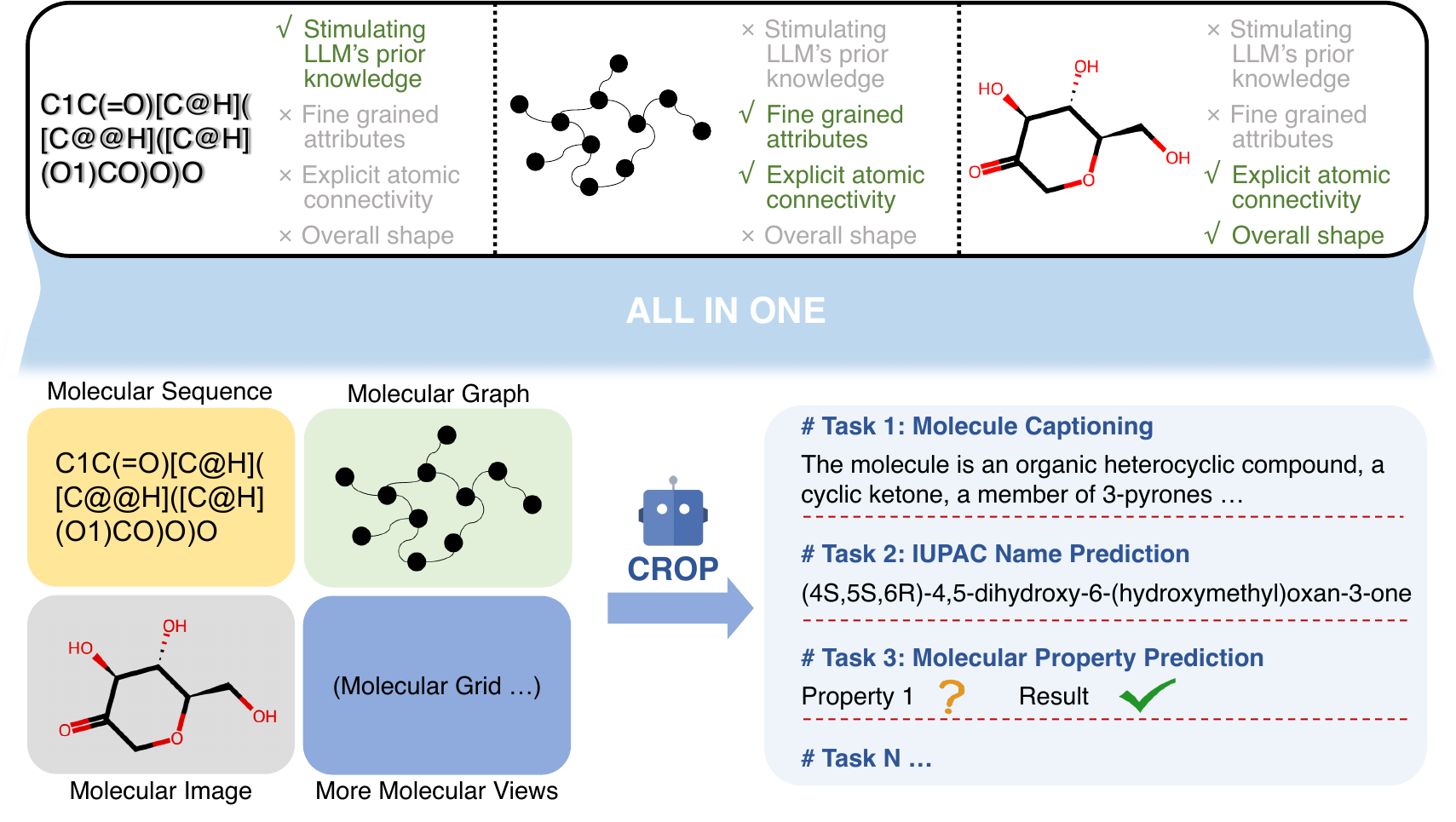}
  \vspace{-15pt}
  \caption{
    An overview of the strengths and weaknesses of each molecular view. Taking that into account, \modelname{} comprehensively utilizes diverse views to enhance molecular understanding capabilities.
  }
  \vspace{-14pt}
  \label{fig:page_begin}
\end{figure}

To address that, recent works~\citep{su2022molecular,liu2023multi,liu2023molca,cao2023instructmol} have preliminarily explored integrating graph-based representations into LLMs, where molecules are modeled as graphs, with atoms as nodes and chemical bonds as edges. While these graph representations effectively capture \textit{topological} relationships between atoms~\citep{li2022deep}, they still exhibit limitations. Specifically, graph representations only encode topological relationships, allowing a single graph to correspond to infinite variety of node arrangements. This ambiguity makes it challenging to represent critical characteristics such as molecular spatial configuration and overall shape, as illustrated in Figure~\ref{fig:page_begin}. Furthermore, when processing complex molecular graphs, graph-based approaches frequently encounter issues such as over-smoothing and over-squashing~\citep{chen2020measuring,keriven2022not}, impeding the effective utilization of graph representations by LLMs.

As a typical representation of the molecular \textit{spatial} view, molecular images provide complementary information about the spatial configuration and overall shape of molecules.
This visual representation naturally encodes important structural features that are difficult to derive from graphs, 
such as symmetry planes, functional group positions and rigidity of molecules~\citep{yi2022micer}. 
Benefiting from sustained advances in computer vision, some studies~\citep{xiang2023chemical,DenseNet121,2DConvNet} that utilize images for molecular modeling have achieved impressive results in discriminative tasks, including molecular property prediction and drug target identification.
Specifically, ImageMol~\citep{zeng2022accurate} conducted 5 carefully crafted pretraining tasks on 10 million molecular images, outperforming existing sequence-based and graph-based methods on 10 regression tasks of GPCRs (G protein-coupled receptors) and 10 classification tasks of kinases in compound-protein binding prediction. 
While molecular images have demonstrated success in discriminative tasks, their application to generative tasks with LLMs remains largely unexplored. This work pioneers the integration of molecular visual representations to enhance LLM performance in generative tasks, such as molecule captioning and IUPAC name prediction.

Given the complementary nature of topological and spatial structures of molecules, works~\citep{liu2023molca,cao2023instructmol} that solely introduce graph views still fundamentally limit LLMs' molecular understanding capabilities. Our key insight is that integrating topological relationships from graphs and spatial configurations from images enables a comprehensive understanding of molecular structures.

Directly concatenating embeddings from graph and image views as input would result in excessive consumption of the LLM's limited context length, and this issue exacerbates as more views are introduced, as shown in Figure~\ref{fig:arch_compare} (\textit{arch1}). 
In fact, there is substantial information irrelevance and overlap among these embeddings. For instance, a significant portion of the image embeddings correspond to the blank areas in the molecular images, and all of these molecular views capture the overlapping information of atomic and bond types.
Therefore, efficiently resampling key structural features from these molecular views becomes crucial.
While independent resampling the graph and image embeddings can reduce input length (Figure~\ref{fig:arch_compare}, \textit{arch2}), it still faces increased consumption of context length when accommodating additional views.
Besides, the lack of guidance from chemical domain knowledge limits the effectiveness of resampling.
Therefore, as shown in Figure~\ref{fig:arch_compare} (\textit{arch3}), we propose to jointly resample multiple structural views into fixed-length cross-view prefixes for LLMs. To boost the quality of the generated prefixes, we utilize the LLM’s self-encoded SMILES as the resampling guidance, which are enriched with the LLM's prior chemical knowledge~\citep{liu2024git}.

To this end, we propose \modelname{}, an MLLM that demonstrates outstanding molecular understanding capabilities, augmented with multiple structural views, as illustrated in Figure~\ref{fig:framework}. \modelname{} partitions the LLM backbone into lower and upper segments. The whole forward propagation process is conducted as follows: (\textit{1}) the LLM's lower segment processes SMILES strings to generate chemical knowledge-aware guidance, referred to as SMILES guidance;
(\textit{2}) the SMILES Guided Resampler adopts the SMILES guidance to resample molecular graphs and images jointly; (\textit{3}) the Structural Embedding Gate converts the derived structural embeddings into fixed-length cross-view prefixes; (\textit{4}) the LLM's upper segment processes both SMILES and prefixes to obtain a comprehensive understanding of molecules.
This architecture substantially enhances the effectiveness of the resampling process. Meanwhile, injecting prefixes into multiple layers of the LLM allows for deep interaction with molecular structural information, facilitating a more accurate understanding of molecular structures.

In summary, our contributions are as follows:

\begin{itemize}

\item We identify the fundamental limitations of current molecular MLLMs that rely solely on the graph view, which captures only topological relationships. In this work, we propose leveraging the complementary topological and spatial information conveyed by molecular graph and image views to jointly advance the molecular understanding capabilities of LLMs.

\item We propose \modelname{}, an innovative and scalable MLLM architecture that can accommodate multiple structural views to jointly enhance molecular understanding while maintaining computational efficiency.

\item Through extensive evaluation, we demonstrate \modelname{} achi-eves significant performance gains across a wide range of tasks, including molecule captioning, IUPAC name prediction, and molecular property prediction, highlighting the superiority of our multi-view integration approach.

\end{itemize}

%
%
\section{Related Work}

\subsection{Molecule Modeling}
SMILES and SELFIES strings can be modeled by language models in a manner similar to text sequences. Models like KV-PLM~\citep{zeng2022deep}, MolT5~\citep{edwards2022translation} and Galactica~\citep{taylor2022galactica} excel in molecule-related tasks by bridging SMILES and biomedical text. 
In molecular graph modeling, both Graph Neural Networks (GNNs)~\citep{scarselli2008graph} and Graph Transformers~\citep{ying2021transformers} are widely employed. Pretraining is conducted at both the graph and node levels to capture the global and local information~\citep{zhu2021pre,rong2020self,velickovic2019deep}.
Molecular images offer distinct advantages in depicting the spatial configuration and overall shape of molecules. These images can be rendered by RDKit~\citep{bento2020open} or captured using physical microscopy~\citep{physical}. Chemception~\citep{chemception}, 2DConvNet~\citep{2DConvNet} and DenseNet121~\citep{DenseNet121} are pioneering works utilizing molecular images for predicting chemical properties, compound toxicity, and contaminant reactivity respectively, demonstrating the potential of molecular images for promoting downstream tasks. Recently, ImageMol~\citep{zeng2022accurate} conducts massive self-supervised pretraining on 10 million unlabeled molecules, outperforming sequence-based and graph-based models across various benchmarks. 
Despite the success of molecular images in discriminative tasks, their potential to enhance the performance of LLMs in generative tasks remains largely unexplored.

\subsection{Multimodal Large Language Models}
MLLMs are capable of processing various modalities beyond text, most of which are tailored for natural modalities such as image~\citep{dockylin} and audio~\citep{xu2024secap,audiogpt}.
However, MLLMs designed for specialized modalities such as molecular graphs, images, and grids~\citep{xie2020multitask} have not been sufficiently explored, which inspires us to propose \modelname{}, a specialized MLLM focusing on molecular modalities in the chemical field.
BLIP-2~\citep{li2023blip} and LLaVA~\citep{liu2023visual} are two representative MLLM architectures, utilizing compressed and uncompressed multimodal embeddings, respectively.
Considering the limited context length of LLMs and substantial information overlap among molecular views, we propose to derive efficient cross-view prefixes for LLMs, with LLM's prior chemical knowledge as resampling guidance.

\subsection{MLLMs for Molecular Science}
In the field of chemistry, beyond the commonly-used molecular sequence view, DrugChat~\citep{liang2023drugchat}, InstructMol~\citep{cao2023instructmol}, MolTC~\citep{fang2024moltc}, GIT-Mol~\citep{liu2024git} and MolCA~\citep{liu2023molca} introduce the graph view additionally to advance molecular understanding of LLMs.
However, molecules inherently exhibit various structural views, and each of them exhibits distinct strengths and weaknesses. The model's performance remains limited when relying solely on a single graph view.
In this research, we propose \modelname{} to integrate the strengths of both graph and image views to collaboratively advance the molecular understanding capabilities.  
Moreover, our architecture enables seamless incorporation of additional structural views while maintaining computational efficiency.

\begin{figure}[t!]
  \centering
  \includegraphics[width=\linewidth]{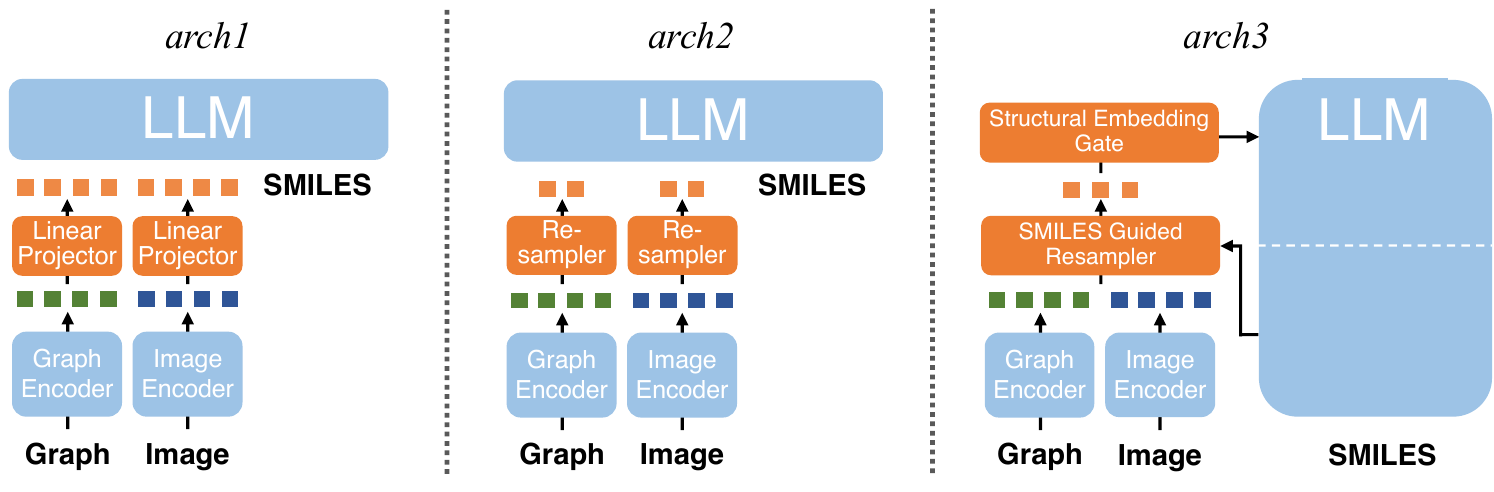}
  \vspace{-15pt}
  \caption{
    Comparison of three MLLM architectures with different ways to process embeddings from multiple molecular views. \modelname{} (\textit{arch3}) jointly resample graph and image views into fixed-length cross-view prefixes for LLMs, with LLM’s prior chemical knowledge as resampling guidance, possessing both efficiency and effectiveness.
  }
  \vspace{-9pt}
  \label{fig:arch_compare}
\end{figure}

\begin{figure*}[t!]
  \centering
  \includegraphics[width=\linewidth]{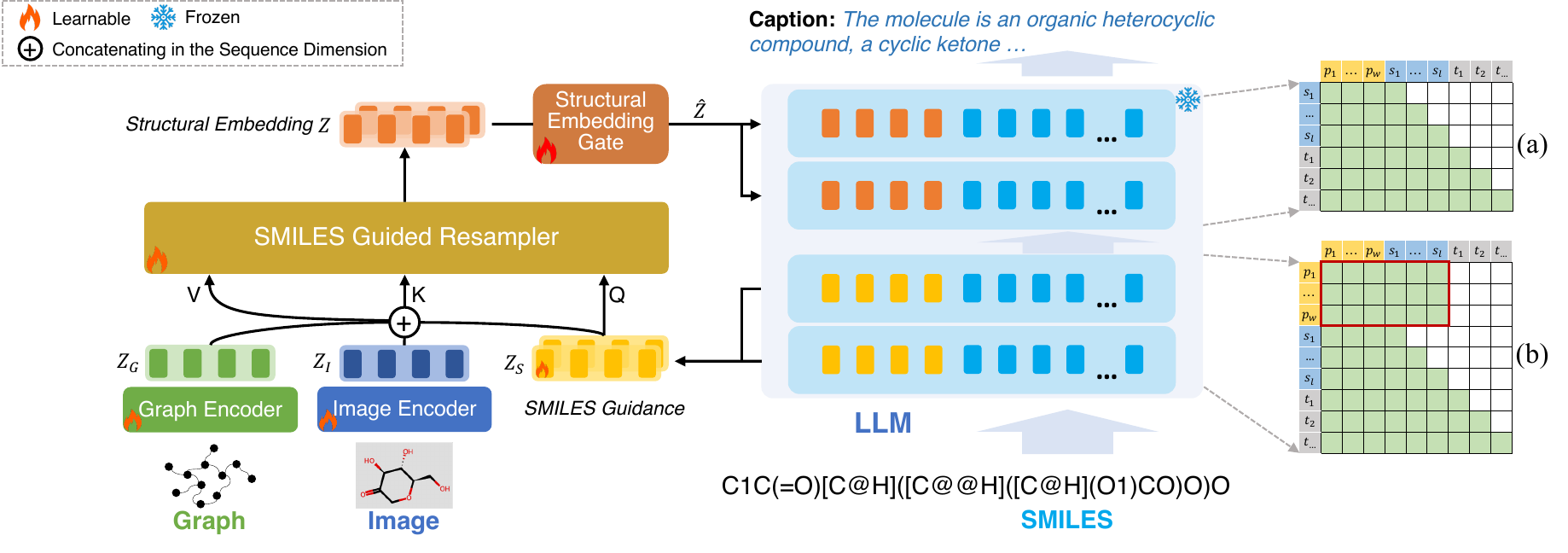}
  \vspace{-20pt}
  \caption{
  The architecture of \modelname{}. The SMILES Guided Resampler utilizes SMILES guidance $Z_S$ derived from the LLM's lower segment to guide the resampling of graph and image views. The Structural Embedding Gate converts the derived structural embeddings $Z$ into prefixes $\hat{Z}$ for the LLM's upper segment. (a) The standard attention mask of the LLM when handling prefixes, where $p$, $s$ and $t$ denote prefix, SMILES and plain text respectively. This is used in the LLM's upper segment. (b) The modified attention mask enabling prefixes to perceive SMILES tokens behind. This is employed in the LLM's lower segment.
  }
  \label{fig:framework}
\end{figure*}

%
%

\section{Methodology}
\subsection{Problem Definition}
In this work, we adopt SMILES as the sequence view to leverage the prior knowledge of LLMs, molecular graphs as the topological structure view to capture atomic 
connections, and molecular images as the spatial structure view to encode molecular configurations.
Let $\boldsymbol{S}=(s_1, s_2, \dots, s_l)$ be a molecule SMILES string tokenized based on characters, where $l$ is the number of characters. Let $\boldsymbol{G}=(V, E)$ be a molecule graph, where $V=\{v_1, v_2, \dots, v_n\}$ is the set of $n$ graph nodes and $E$ is the set of graph edges. Let $\boldsymbol{I}$ be a molecule image. Given a molecule's $\boldsymbol{S}$, $\boldsymbol{G}$ and $\boldsymbol{I}$, generative tasks such as molecule captioning and IUPAC name prediction require generating a corresponding text $y$. Classification tasks such as molecule property prediction require generating a probability distribution $p$, with cross-entropy loss used for optimization. 

\subsection{Model Architecture}
\paragraph{\textbf{Overview}}
As illustrated in Figure~\ref{fig:framework}, \modelname{} comprises four primary components: the LLM, molecule encoders, the SMILES Guided Resampler and the Structural Embedding Gate.
The LLM is partitioned into lower and upper segments.
The lower segment processes the SMILES to derive the SMILES guidance $Z_S$.
Then, the SMILES Guided Resampler utilizes the SMILES guidance $Z_S$ to jointly resample molecular graph embeddings $Z_G$ and image embeddings $Z_I$ derived from respective encoders, and output the structural embeddings $Z$.
After that, the Structural Embedding Gate converts the structural embeddings $Z$ into fixed-length cross-view prefixes $\hat{Z}$.
Finally, the LLM's upper segment processes both the SMILES hidden states from the lower segment and prefixes $\hat{Z}$ to achieve a comprehensive understanding of molecules. 
Benefiting from the SMILES guidance $Z_S$, which is enriched with the LLM’s prior chemical knowledge, the resampling possesses high effectiveness. Moreover, injecting cross-view prefixes $\hat{Z}$ into multiple layers of the LLM enables deep interaction with molecular structural information.

\paragraph{\textbf{Molecule Encoders}}
The molecular graph encoder consists of five Graph Isomorphism Network (GIN) layers \citep{xu2018powerful} initialized from moleculeSTM \citep{liu2023multi}. Molecular graphs are encoded into representations $Z_G \in \mathbb{R}^{n \times d}$, where each node representation contains local structural information of the neighboring subgraph. The molecular image encoder adopts the pretrained ResNet18 from ImageMol~\citep{zeng2022accurate}. The original molecular image is encoded to a feature map with dimension $H \times W \times d$. This feature map is then flattened to $Z_I \in \mathbb{R}^{p \times d}$, where $p = HW$. These process can be formalized as:
\begin{equation}
\begin{aligned}
Z_G &= GraphEncoder(\boldsymbol{G}), \\
Z_I &= ImageEncoder(\boldsymbol{I}).
\end{aligned}
\end{equation}

\paragraph{\textbf{SMILES Guidance (SG)}}
We adopt Galactica~\citep{taylor2022galactica} as the LLM backbone of \modelname{}, which is pretrained on an extensive chemical corpus and known for its strong proficiency in chemistry.
Benefiting from this, we adopt the SMILES representations within Galactica as guidance, which are enriched with Galactica's prior chemical knowledge, to facilitate resampling key structural features from graphs and images.
To obtain the SMILES guidance $Z_S$ from the LLM and return cross-view prefixes $\hat{Z}$ to the LLM during a single forward propagation, we partition the LLM into lower and upper segments, consisting of $b$ and $u$ layers respectively.

We prepend $w$ learnable vectors to all layers in the lower segment, enabling extensive interaction with SMILES hidden states via the LLM's attention layers. These fixed-length vectors then serve as the SMILES guidance $Z_S \in \mathbb{R}^{b \times w \times d}$.
It is worth noting that the prepended vectors cannot directly perceive the SMILES tokens behind, due to the causal attention mechanism within the LLM.
As illustrated in Figure~\ref{fig:framework} (a), with prepended vectors, the standard attention calculation process can be formalized as:
\begin{equation}
\begin{split}
&\mathrm{AttentionLayer}(H,H_p,H_p) \\
= &\mathrm{softmax}\left(\frac{(HW_Q)(H_pW_K)^T}{\sqrt{d_k}} + M\right)(H_pW_V),
\end{split}
\end{equation}
where $H \in \mathbb{R}^{l \times d}$ denotes SMILES hidden states, and $H_p \in \mathbb{R}^{(w+l) \times d}$ represents the concatenation of the prepended vectors and SMILES hidden states. $W_Q, W_K \text{ and } W_V \in \mathbb{R}^{d \times d}$ are the query, key and value transform matrices respectively. $M$ is the triangular causal attention mask, with $M_{i,j \leq i+w}=0$ and $M_{i,j > i+w}=-\infty$.  

In order to enable the prepended vectors to perceive the SMILES tokens behind, as illustrated in Figure~\ref{fig:framework} (b), we modify the standard attention calculation process in the lower segment, which can be formalized as:
\begin{equation}
\begin{split}
&\mathrm{AttentionLayer}(H_p,H_p,H_p) \\
= &\mathrm{softmax}\left(\frac{(H_pW_Q)(H_pW_K)^T}{\sqrt{d_k}} + M'\right)(H_pW_V).
\end{split}
\end{equation}
where $M'$ denotes the modified attention mask, with $M'_{i \leq w, j \leq w+l}=0$, $M'_{i \leq w, j > w+l}=-\infty$, $M'_{i > w, j \leq i}=0$ and $M'_{i > w, j > i}=-\infty$.

\paragraph{\textbf{SMILES Guided Resampler (SGR)}}
By leveraging the SMILES guidance, SGR proceeds to jointly resample lengthy graph and image embeddings into fixed-length structural embeddings, which integrate the strengths of both molecular views. 

SGR consists of multiple transformer layers. 
Graph embeddings $Z_G$, image embeddings $Z_I$ and SMILES guidance $Z_S$ are concatenated along the sequence dimension to serve as keys and values. The SMILES guidance $Z_S$, enriched with LLM's prior chemical knowledge, serve as queries:
\begin{gather}
Keys, Values = [Z_G, Z_I, Z_S] \in \mathbb{R}^{b \times (n+p+w) \times d}, \\
Queries = Z_S \in \mathbb{R}^{b \times w \times d}, \\
Z = SGR(Queries, Keys, Values) \in \mathbb{R}^{b \times w \times d}.
\end{gather}
where the resampling process is conducted through the cross attention mechanism in SGR's transformer layers, producing compact structural embeddings $Z$.

\begin{figure}[b!]
  \centering
  \includegraphics[width=0.8\linewidth]{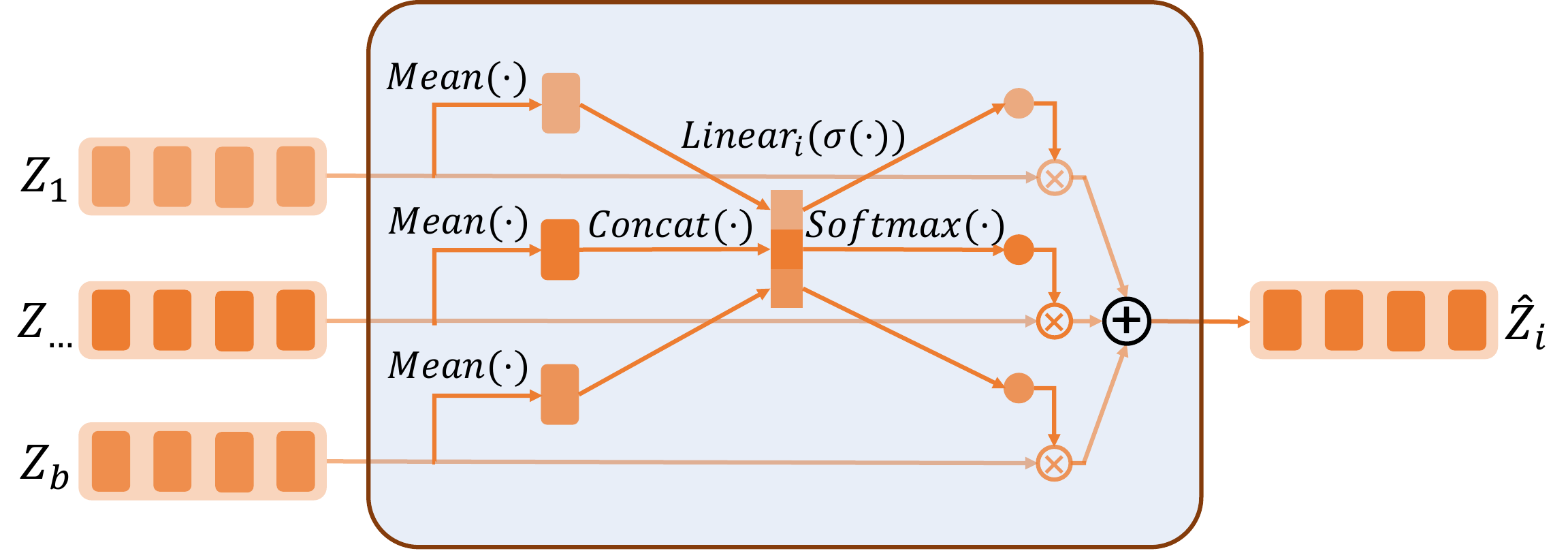}
  \caption{
  The architecture of the Structural Embedding Gate for the LLM's $\mathrm{Layer}_i$, which sums $b$ groups of $Z_j \in \mathbb{R}^{w \times d}$ with the weighted vector $P_i \in \mathbb{R}^{b}$ to obtain the prefixes $\hat{Z}_i \in \mathbb{R}^{w \times d}$.
  }
  \label{fig:gate_module}
\end{figure}

Beyond molecular graph and image views, SGR can further accommodate more structural views by simply concatenating their embeddings with $[Z_G, Z_I, Z_S]$, and serve as keys and values. This design ensures the extensibility and flexibility of SGR.

\begin{table*}[t!]
    \centering
    \setlength{\tabcolsep}{3.6pt}
    \caption{Molecule captioning results on PubChem324k and CheBI-20 datasets. \textbf{Bold} denotes the best performance.}
    \begin{tabular}{l|lll|cccccc} \toprule
        Dataset & Model & Modalities & TrainableParams & BLEU-2 & BLEU-4 & ROUGE-1 & ROUGE-2 & ROUGE-L & METEOR \\ 
        \midrule
        \multirow{10}{*}{PubChem324k} & GPT-4o & $\boldsymbol{S}$ & \quad-\quad, 3-shot & 16.5 & 7.1 & 30.8 & 11.3 & 23.1 & 22.5 \\
        & $\text{Llama3}_\textit{Instruct8B}$ & $\boldsymbol{S}$ & \quad-\quad, 3-shot & 14.4 & 6.4 & 30.3 & 12.8 & 24.7 & 20.3 \\
        & BioT5 & $\boldsymbol{S}$ & 252M, full ft & 42.9 & 34.3 & 53.1 & 39.8 & 47.5 & 48.7 \\
        & MolT5-Large & $\boldsymbol{S}$ & 780M, full ft & 30.2 & 22.2 & 41.5 & 25.9 & 34.8 & 36.6 \\
        & MoMu-Large & $\boldsymbol{S}+\boldsymbol{G}$ & 782M, full ft & 31.1 & 22.8 & 41.8 & 25.7 & 36.7 & 36.2 \\
        & $\text{MolCA}_\textit{Galac1.3B}$ & $\boldsymbol{S}+\boldsymbol{G}$ & 100M, LoRA ft & 38.7 & 30.3 & 50.2 & 35.9 & 44.5 & 45.6 \\
        \cmidrule(lr){2-10}
        & $\text{\modelname}_{\textit{Galac1.3B}}$ & $\boldsymbol{S}$ & 71M, LoRA ft & 36.6 & 29.1 & 48.7 & 34.8 & 43.1 & 43.5 \\
        & $\text{\modelname}_{\textit{Galac1.3B}}$ & $\boldsymbol{S}+\boldsymbol{G}$ & 71M, LoRA ft & 43.4 & 35.4 & 54.0 & 40.1 & 48.7 & 49.5 \\
        & $\text{\modelname}_{\textit{Galac1.3B}}$ & $\boldsymbol{S}+\boldsymbol{I}$ & 71M, LoRA ft & 43.1 & 34.9 & 53.5 & 39.7 & 48.3 & 49.1 \\
        & $\text{\modelname}_{\textit{Galac1.3B}}$ & $\boldsymbol{S}+\boldsymbol{G}+\boldsymbol{I}$ & 71M, LoRA ft & \textbf{44.9} & \textbf{36.7} & \textbf{54.8} & \textbf{41.1} & \textbf{49.5} & \textbf{50.8} \\
        \addlinespace
        \bottomrule
        \addlinespace
        \multirow{13}{*}{CheBI-20} & GPT-4o & $\boldsymbol{S}$ & \quad-\quad, 3-shot & 21.9 & 9.8 & 35.9 & 13.6 & 26.7 & 27.6 \\
        & $\text{Llama3}_\textit{Instruct8B}$ & $\boldsymbol{S}$ & \quad-\quad, 3-shot & 20.9 & 9.4 & 35.8 & 15.5 & 28.9 & 25.4 \\
        & $\text{MolReGPT}_\textit{GPT4}$ & $\boldsymbol{S}$ & RAG, 10-shot & 60.7 & 52.5 & 63.4 & 47.6 & 56.2 & 61.0 \\ 
        & MolXPT & $\boldsymbol{S}$ & 350M, full ft & 59.4 & 50.5 & 66.0 & 51.1 & 59.7 & 62.6 \\
        & BioT5 & $\boldsymbol{S}$ & 252M, full ft & 63.5 & 55.6 & 69.2 & 55.9 & 63.3 & 65.6 \\
        & MoMu-Large & $\boldsymbol{S}+\boldsymbol{G}$ & 782M, full ft & 59.9 & 51.5 & - & - & 59.3 & 59.7 \\
        & GIT-Mol & $\boldsymbol{S}+\boldsymbol{G}$ & 210M, LoRA ft & 35.2 & 26.3 & 57.5 & 48.5 & 56.0 & 53.3 \\
        & InstructMol & $\boldsymbol{S}+\boldsymbol{G}$ & \quad-\quad, LoRA ft & 47.5 & 37.1 & 56.6 & 39.4 & 50.2 & 50.9 \\
        & $\text{MolCA}_\textit{Galac1.3B}$ & $\boldsymbol{S}+\boldsymbol{G}$ & 110M, LoRA ft & 62.0 & 53.1 & 68.1 & 53.7 & 61.8 & 65.1 \\
        \cmidrule(lr){2-10}
        & $\text{\modelname}_{\textit{Galac1.3B}}$ & $\boldsymbol{S}$ & 71M, LoRA ft & 58.4 & 49.6 & 67.0 & 51.3 & 60.3 & 62.7 \\
        & $\text{\modelname}_{\textit{Galac1.3B}}$ & $\boldsymbol{S}+\boldsymbol{G}$ & 71M, LoRA ft & 63.8 & 55.8 & 69.0 & 55.0 & 63.5 & 66.1 \\
        & $\text{\modelname}_{\textit{Galac1.3B}}$ & $\boldsymbol{S}+\boldsymbol{I}$ & 71M, LoRA ft & 62.8 & 54.2 & 68.9 & 54.7 & 62.8 & 65.8 \\
        & $\text{\modelname}_{\textit{Galac1.3B}}$ & $\boldsymbol{S}+\boldsymbol{G}+\boldsymbol{I}$ & 71M, LoRA ft & \textbf{64.6} & \textbf{56.2} & \textbf{69.8} & \textbf{55.9} & \textbf{63.9} & \textbf{66.7} \\
        \bottomrule
    \end{tabular}
    \label{table:mol_caption}
\end{table*}

\paragraph{\textbf{Structural Embedding Gate (SEG)}}
Considering the lower and upper segment of the LLM may contain different number of layers, namely $b \neq u$, we propose the SEG.
SEG can flexibly convert structural embeddings $Z \in R^{b \times w \times d}$ into fixed-length cross-view prefixes $\hat{Z} \in R^{u \times w \times d}$, which are then prepended to the LLM's upper segment. These prefixes empower the LLM to understand molecular structures accurately and comprehensively.

Specifically, we sum $b$ groups of $Z_j \in \mathbb{R}^{w \times d}$ with the weighted vector $P_i \in \mathbb{R}^{b}$ to obtain the prefixes $\hat{Z}_i \in \mathbb{R}^{w \times d}$ for $\mathrm{Layer}_i$ in the LLM's upper segment, as shown in Figure~\ref{fig:gate_module}. The calculation is formalized as follows:
\begin{gather}
V = \mathrm{Concat}(\mathrm{Mean}(Z_1);\dots;\mathrm{Mean}(Z_b)), \\
P_i = \mathrm{Softmax}(\mathrm{Linear}_i({\sigma(V)})), \\
\hat{Z}_i = P_iZ.
\end{gather}
where $Z_{j=1,2,\dots,b} \in \mathbb{R}^{w \times d}$ indicates the $j$th group of embedding in $Z$. $\mathrm{Mean}(\cdot)$ denotes mean pooling along the sequence dimension, yielding $\mathrm{Mean}(Z_j) \in \mathbb{R}^d$. $\mathrm{Concat}(\cdot)$ concatenates embeddings along the hidden dimension, producing $V \in \mathbb{R}^{bd}$. $\sigma(\cdot)$ is the activation function. $\mathrm{Linear}_i(\cdot)$ has an input dimension of $bd$ and an output dimension of $b$, thus $P_i \in \mathbb{R}^b$.

SEG can flexibly convert any $b$ groups of structural embeddings into specified $u$ groups of cross-view prefixes, thus the LLM can be partitioned arbitrarily, where the lower segment is unimodal and the upper segment is multimodal.

\subsection{Training Strategies}
As Galactica is primarily pretrained on SMILES, we adopt SMILES as the molecular sequence view in this research to fully stimulate Galactica's prior chemical knowledge.
The training process consists of two stages.
In the pretraining stage (Stage 1), \modelname{} performs the molecule captioning task conditioned on molecular SMILES, graphs, and images.
Molecular graphs and images could be obtained using the RDKit toolkit according to SMILES. 
Common data augmentation techniques for images are applied to enhance the LLM's ability to leverage the structural information in molecular images.
The primary objective of this stage is to establish initial alignment between multiple views and the LLM. Therefore, we freeze the molecular graph encoder, image encoder and the LLM, and focus on training the bridging modules, including the SMILES guidance, SGR and SEG. In the fine-tuning stage (Stage 2), to pursue optimal performance on downstream tasks, in addition to the aforementioned bridging modules, we unfreeze the molecular graph encoder and image encoder and utilize LoRA~\citep{hu2021lora} to fine-tune the LLM.

\begin{table*}[tb!]
    \centering
    \setlength{\tabcolsep}{3.5pt}
    \caption{IUPAC name prediction results on PubChem324k (Baseline results are from~\citep{liu2023molca}).}
    \begin{tabular}{lll|cccccc} \toprule
    Model & Modalities & TrainableParams & BLEU-2 & BLEU-4 & ROUGE-1 & ROUGE-2 & ROUGE-L & METEOR \\ 
    \midrule
    GPT-4o~\citep{openai_hello_gpt_4o} & $\boldsymbol{S}$ & \quad-\quad, 3-shot & 42.6 & 26.1 & 40.5 & 13.4 & 32.6 & 41.1 \\
    $\text{Llama3}_\textit{Instruct8B}$~\citep{dubey2024llama} & $\boldsymbol{S}$ & \quad-\quad, 3-shot & 31.9 & 16.0 & 28.9 & 5.0 & 22.3 & 26.8 \\
    BioT5~\citep{biot5} & $\boldsymbol{S}$ & 252M, full ft & 79.4 & 72.6 & 75.4 & 55.7 & 69.5 & 75.8 \\
    GIT-Mol~\citep{liu2024git} & $\boldsymbol{S}+\boldsymbol{G}$ & 210M, LoRA ft & 58.3 & 51.7 & 54.5 & 32.6 & 50.2 & 55.7 \\
    $\text{MolCA}_\textit{Galac1.3B}$~\citep{liu2023molca} & $\boldsymbol{S}+\boldsymbol{G}$ & 100M, LoRA ft & 75.0 & 66.6 & 69.6 & 48.2 & 63.4 & 72.1 \\
    \midrule
    $\text{\modelname}_{\textit{Galac1.3B}}$ & $\boldsymbol{S}$ & 71M, LoRA ft & 74.5 & 65.9 & 68.7 & 47.3 & 62.5 & 71.4 \\
    $\text{\modelname}_{\textit{Galac1.3B}}$ & $\boldsymbol{S}+\boldsymbol{G}$ & 71M, LoRA ft & 80.8 & 73.2 & 77.5 & 57.5 & 72.0 & 78.2 \\
    $\text{\modelname}_{\textit{Galac1.3B}}$ & $\boldsymbol{S}+\boldsymbol{I}$ & 71M, LoRA ft & 80.6 & 72.7 & 77.1 & 56.9 & 71.4 & 77.9 \\
    $\text{\modelname}_{\textit{Galac1.3B}}$ & $\boldsymbol{S}+\boldsymbol{G}+\boldsymbol{I}$ & 71M, LoRA ft & \textbf{81.5} & \textbf{74.3} & \textbf{78.5} & \textbf{58.6} & \textbf{72.9} & \textbf{78.8} \\
    \bottomrule
    \end{tabular}
    \label{table:mol_iupac}
\end{table*}

\begin{table*}[t!]
    \centering
    \caption{Molecule property prediction results on 6 datasets in MoleculeNet. The scaffold splits~\citep{yang2019analyzing} are adopted. Baseline results are from their original papers. ROC-AUC scores are calculated across 5 random seeds.}
    \begin{tabular}{ll|cccccc|c} \toprule
    Model & Modalities & Tox21 ↑ & \ ToxCast ↑ & \ Sider ↑ & ClinTox ↑ & BBBP ↑ & Bace ↑ & Mean \\
    \midrule
    KV-PLM~\citep{zeng2022deep} & $\boldsymbol{S}$ & 72.1$\pm$1.0 & 55.0$\pm$1.7 & 59.8$\pm$0.6 & - & 70.5$\pm$0.5 & 78.5$\pm$2.7 & 67.2  \\
    Mole-BERT~\citep{xia2022mole} & $\boldsymbol{G}$ & 76.8$\pm$0.5 & 64.3$\pm$0.2 & 62.8$\pm$1.1 & 78.9$\pm$3.0 & 71.9$\pm$1.6 & 80.8$\pm$1.4 &72.6 \\
    MoMu~\citep{su2022molecular} & $\boldsymbol{G}$ & 75.6$\pm$0.3 & 63.4$\pm$0.5 & 60.5$\pm$0.9 & 79.9$\pm$4.1 & 70.5$\pm$2.0 & 76.7$\pm$2.1 & 71.1 \\
    GIT-Mol~\citep{liu2024git} & $\boldsymbol{S}+\boldsymbol{G}$ & 75.9$\pm$0.5 & \textbf{66.8$\pm$0.5} & 63.4$\pm$0.8 & 88.3$\pm$1.2 & \textbf{73.9$\pm$0.6} & 81.1$\pm$1.5 & 74.9 \\
    $\text{MolCA}_\textit{Galac1.3B}$~\citep{liu2023molca} & $\boldsymbol{S}+\boldsymbol{G}$ & \underline{77.2$\pm$0.5} & \underline{64.5$\pm$0.8} & 63.0$\pm$1.7 & 89.5$\pm$0.7 & 70.0$\pm$0.5 & 79.8$\pm$0.5 & 74.0 \\
    \midrule
    $\text{\modelname}_{\textit{Galac1.3B}}$ & $\boldsymbol{S}$ & 72.6$\pm$0.6 & 58.3$\pm$0.7 & 63.3$\pm$1.5 & 90.8$\pm$1.8 & 71.0$\pm$1.5 & 81.0$\pm$1.2 & 72.8  \\
    $\text{\modelname}_{\textit{Galac1.3B}}$ & $\boldsymbol{S}+\boldsymbol{G}$ & 76.2$\pm$0.4 & 60.9$\pm$0.6 & \underline{65.1$\pm$0.7} & \underline{93.7$\pm$0.9} & 71.2$\pm$1.0 & \underline{83.5$\pm$0.5} & \underline{75.1}  \\
    $\text{\modelname}_{\textit{Galac1.3B}}$ & $\boldsymbol{S}+\boldsymbol{I}$ & 75.4$\pm$0.6 & 60.7$\pm$0.4 & 66.2$\pm$1.2 & 92.0$\pm$1.4 & 72.2$\pm$0.8 & 82.9$\pm$0.8 & 74.9  \\
    $\text{\modelname}_{\textit{Galac1.3B}}$ & $\boldsymbol{S}+\boldsymbol{G}+\boldsymbol{I}$ & \textbf{77.5$\pm$0.2} & 61.4$\pm$0.4 & \textbf{67.3$\pm$0.7} & \textbf{94.6$\pm$0.6} & \underline{72.6$\pm$0.6} & \textbf{84.2$\pm$0.7} & \textbf{76.3}  \\
    \bottomrule
    \end{tabular}
    \label{table:mol_mpp}
\end{table*}

To conduct a fair comparison with baselines and to analyze the contributions of molecular graphs and images respectively, we develop four variants of \modelname{} by selectively including different structural representations alongside SMILES. Specifically, by masking graph embeddings $Z_G$ or image embeddings $Z_I$, we develop $\text{\modelname}_{(\boldsymbol{S}+\boldsymbol{G}+\boldsymbol{I})}$, $\text{\modelname}_{(\boldsymbol{S}+\boldsymbol{G})}$ and $\text{\modelname}_{(\boldsymbol{S}+\boldsymbol{I})}$. When all structural views are excluded, $\text{\modelname}_{(\boldsymbol{S})}$ reduces to the original Galactica operating only on SMILES input.

%
%

\section{Experiments}
\subsection{Experimental Setup}
\paragraph{\textbf{Datasets}}
We pretrain CROP on PubChem324k's pretraining subset~\citep{liu2023molca}, which contains about 300k molecule-text pairs of relatively low-quality.
For the molecule captioning task, we evaluate \modelname{}'s performance on the standard PubChem324k and CheBI-20~\citep{edwards2022translation} datasets.
For the IUPAC name prediction task, we evaluate \modelname{}'s performance on the standard PubChem324k dataset.
For the molecule property prediction task, we evaluate \modelname{}'s performance on various sub-datasets in MoleculeNet~\citep{wu2018moleculenet}: Tox21, ToxCast, Sider, ClinTox, BBBP, and Bace, with the scaffold splits~\citep{yang2019analyzing} adopted.

\paragraph{\textbf{Implementation Details}}
\modelname{} is pretrained for 20 epochs and finetuned for 100 epochs on all downstream task datasets respectively. The best-performing model on the validation set is selected for testing. To save the context length of the LLM, we experimentally determine the prefix length $w$ as 10.
We analyze the performance of \modelname{} under different partition settings and identify the optimal partition for each task. Specifically, we set $b=12, u=12$ in the molecule captioning and molecule property prediction tasks, and set $b=6, u=18$ in the IUPAC name prediction task.

\begin{figure*}[t!]
    \centering
    \subfloat{\includegraphics[width=0.95\linewidth]{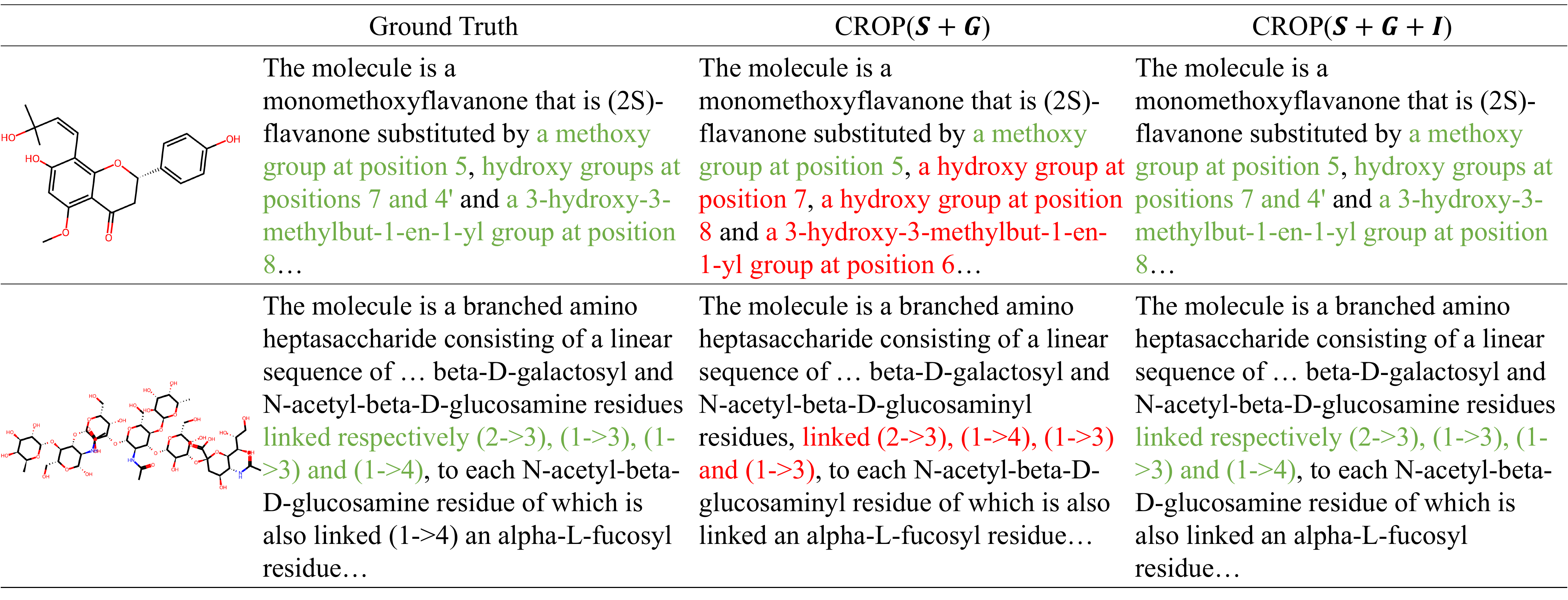}}
    \caption{The captions generated by $\text{\modelname}_{(\boldsymbol{S}+\boldsymbol{G})}$ and $\text{\modelname}_{(\boldsymbol{S}+\boldsymbol{G}+\boldsymbol{I})}$ on example molecules. $\text{\modelname}_{(\boldsymbol{S}+\boldsymbol{G}+\boldsymbol{I})}$ provides more accurate descriptions of substituent positions, types, and the connectivity of branched structures in molecules.}
    \label{fig:caption_sample}
\end{figure*}

\subsection{Evaluation on Downstream Tasks}
\paragraph{\textbf{Molecule Captioning}}
This task requires generating a description about the molecule's properties, structures, biological activity, and other characteristics. In the setting with sequence-only input, leading LLMs such as MolT5~\citep{molt5}, MolXPT~\citep{molxpt} and BioT5~\citep{biot5} are included as baselines.
Besides, we report the performance of GPT-4o and Llama3-Instruct in a few-shot setting to showcase the current progress of mainstream general-purpose LLMs in the field of chemistry. MolReGPT~\citep{MolReGPT} is based on GPT4 with the retrieval-augmented generation (RAG) technique. In the setting with multi-view input, advanced MLLMs such as MoMu~\citep{su2022molecular}, GIT-Mol~\citep{liu2024git}, InstructMol~\citep{cao2023instructmol}, and MolCA~\citep{liu2023molca} are included as baselines. Among them, InstructMol employs the LLaVA architecture~\citep{liu2023visual} and MolCA employs the BLIP-2~\citep{li2023blip} architecture.

The results are reported at Table~\ref{table:mol_caption}. 
Beyond the original SMILES view, when utilizing both graph and image views additionally, $\text{\modelname}_{(\boldsymbol{S}+\boldsymbol{G}+\boldsymbol{I})}$ achieves the best performance.
This highlights the importance of integrating multiple structural views to provide more accurate molecular information, which can mitigate the limitations of relying on a single structural view.
When only adopt the graph view additionally, $\text{\modelname}_{(\boldsymbol{S}+\boldsymbol{G})}$ still outperforms other models by a large margin with fewer trainable parameters, highlighting \modelname{}'s architectural superiority. 
The superiority primarily stems from two aspects. First, \modelname{} introduces SMILES guidance when jointly resampling multiple structural views, thereby effectively leveraging Galactica’s prior chemical knowledge. Second, the generated cross-view prefixes are prepended to all layers in the upper segment of the LLM, enabling it to fully interact with the molecular structural information. In contrast, models such as InstructMol or MolCA provide structural information from the input embedding layer.
Detailed analysis is provided in the Ablation Studies.

\paragraph{\textbf{IUPAC Name Prediction}}
The IUPAC systematic names \citep{favre2013nomenclature} are standardized molecular identifiers that take aspects, such as molecular functional groups and substituents, into account. This task aims at predicting IUPAC name strings from other molecular representations, thus requiring accurately understanding molecular complex structures.
$\text{\modelname}_{(\boldsymbol{S}+\boldsymbol{G}+\boldsymbol{I})}$ outperforms MolCA by 6.5 in BLEU-2 score, as shown in Table~\ref{table:mol_iupac}.

\paragraph{\textbf{Molecule Property Prediction}}
This task involves judging a molecule's potential toxicity and other properties based on its structure. 
Cross-view prefixes and SMILES hidden states from the LLM's last layer are passed through a separate linear head for classification.
As shown in Table~\ref{table:mol_mpp}, on six datasets in MoleculeNet, $\text{\modelname}_{(\boldsymbol{S}+\boldsymbol{G}+\boldsymbol{I})}$ outperforms GIT-Mol by 1.4 ROC-AUC scores on average.

\subsection{Ablation Studies}
\begin{figure}[b!]
    \centering
    \subfloat{\includegraphics[width=\linewidth]{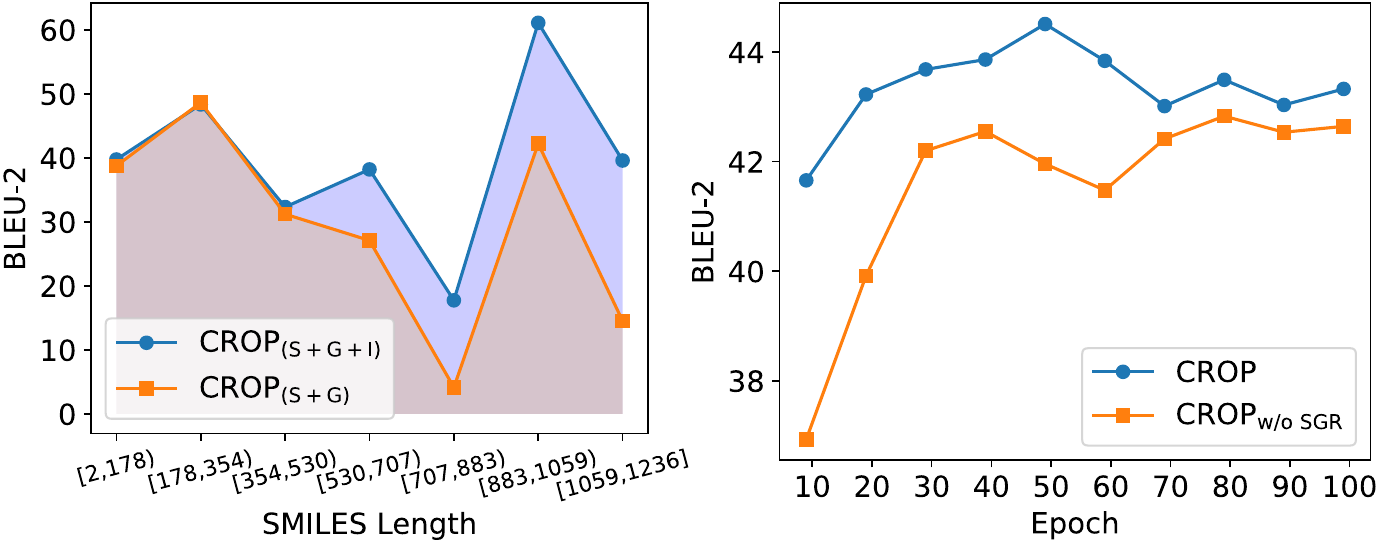}}
    \caption{(Left) The distinct BLEU-2 scores in different SMILES length ranges on the PubChem324k molecule captioning dataset. (Right) The distinct metric curves of \modelname{} and $\text{\modelname}_{\text{w/o \textsc{sgr}\xspace}}$ on the PubChem324k molecule captioning validation set during finetuning.}
    \label{fig:smiles_sgr}
\end{figure}

\paragraph{\textbf{Effectiveness of Integrating Topological and Spatial Structural Views}}
The structures of molecules can be categorized into two types: topological and spatial structures.
However, previous MLLMs primarily focus on the graph view, which emphasizes topological relationships, while neglecting the image view, which excels at capturing molecular spatial configurations. 
In this section, we highlight the benefits of jointly considering the two types of structural information conveyed in molecular graphs and images respectively.
As shown in Table~\ref{table:mol_caption}, compared to $\text{\modelname}_{(\boldsymbol{S})}$, $\text{\modelname}_{(\boldsymbol{S}+\boldsymbol{G})}$ achieves a 6.8 BLEU-2 score improvement on PubChem324k and a 5.4 BLEU-2 score improvement on CheBI-20.
Similarly, $\text{\modelname}_{(\boldsymbol{S}+\boldsymbol{I})}$ improves the BLEU-2 score by 6.5 on PubChem324k and 4.4 on CheBI-20. 
This demonstrates that incorporating either molecular graphs or images independently can enhance the LLM's comprehension of molecular structures.
Furthermore, when both graph and image views are introduced, 
$\text{\modelname}_{(\boldsymbol{S}+\boldsymbol{G}+\boldsymbol{I})}$ outperforms $\text{\modelname}_{(\boldsymbol{S}+\boldsymbol{G})}$ and $\text{\modelname}_{(\boldsymbol{S}+\boldsymbol{I})}$ by a large margin, demonstrating the limitations of relying solely on a single structural view.
For a more detailed analysis, we compare the quality of captions generated by $\text{\modelname}_{(\boldsymbol{S}+\boldsymbol{G})}$ and $\text{\modelname}_{(\boldsymbol{S}+\boldsymbol{G}+\boldsymbol{I})}$ for molecules of different complexity, which is measured by the length of SMILES. As illustrated in Figure~\ref{fig:smiles_sgr} (Left), molecular images enhance the performance of \modelname{} particularly on more complex molecules.

Some examples are provided in Figure~\ref{fig:caption_sample}, where it can be observed that \modelname{}, by incorporating images, is able to more accurately describe the substituent positions and the spatial interconnectivity of units in complex molecules.
Therefore, in this work, we also reveal the effectiveness of molecular images in generative tasks.

\paragraph{\textbf{Effectiveness of SGR}}
The effectiveness of SGR stems from its use of SMILES guidance to guide the resampling process, which is derived from the LLM and enriched with prior chemical knowledge.
To validate the role of SMILES guidance in boosting the quality of the derived cross-view prefixes, we replace SGR with an alternative resampler, in which the SMILES guidance is substituted with learnable vectors which are randomly initialized. This variant is referred to as $\text{\modelname}_{\text{w/o \textsc{sgr}\xspace}}$.
Without the need to obtain SMILES guidance during the forward propagation, the LLM does not need to be partitioned into two segments, allowing us to prepend the cross-view prefixes to all layers of the LLM.
We compare the performance of \modelname{} and $\text{\modelname}_{\text{w/o \textsc{sgr}\xspace}}$ when adopting both graph and image views. As shown in Table~\ref{table:ablations}, \modelname{} consistently outperforms $\text{\modelname}_{\text{w/o \textsc{sgr}\xspace}}$ across diverse tasks. In addition, we also find that \modelname{} converges much faster particularly in the initial stage, and reaches the optimum state earlier than $\text{\modelname}_{\text{w/o \textsc{sgr}\xspace}}$ during the finetuning process, as illustrated in Figure~\ref{fig:smiles_sgr} (Right). These demonstrate the effectiveness of SGR when utilizing SMILES guidance.

\begin{table}[t!]
    \centering
    \setlength{\tabcolsep}{1pt}
    \caption{Ablation results on the molecule captioning, IUPAC name prediction and molecule property prediction tasks. All experiments are conducted with 
    $\boldsymbol{S},\boldsymbol{G},\boldsymbol{I}$ as input simultaneously. The second lines denote $\text{\modelname}_{\text{w/o \textsc{sgr}\xspace}}$ and the third lines denote $\text{\modelname}_{\text{w/o \textsc{seg}\xspace}}$.}
    \begin{tabular}{cc|cccccc}
    \toprule
    \hline
    \rowcolor[gray]{0.85}
    \multicolumn{2}{c|}{\modelname{}} & \multicolumn{6}{c}{Molecule Captioning} \\
    SGR & SEG & $\text{BLEU}_\text{2}$ & $\text{BLEU}_\text{4}$ & $\text{ROUGE}_\text{1}$ & $\text{ROUGE}_\text{2}$ & $\text{ROUGE}_\text{L}$ & METEOR \\
    \midrule
    \ding{51} & \ding{51} & \textbf{44.9} & \textbf{36.7} & \textbf{54.8} & \textbf{41.1} & \textbf{49.5} & \textbf{50.8}\\
    & \ding{51} & 42.4 & 33.5 & 52.4 & 38.9 & 48.1 & 48.1\\
    \ding{51} & & 44.2 & 36.2 & 54.0 & 40.7 & 48.4 & 50.2\\

    \midrule

    \hline
    \rowcolor[gray]{0.85}
    \multicolumn{2}{c|}{\modelname{}} & \multicolumn{6}{c}{IUPAC Name Prediction} \\
    SGR & SEG & $\text{BLEU}_\text{2}$ & $\text{BLEU}_\text{4}$ & $\text{ROUGE}_\text{1}$ & $\text{ROUGE}_\text{2}$ & $\text{ROUGE}_\text{L}$ & METEOR \\
    \midrule
    \ding{51} & \ding{51} & \textbf{81.5} & \textbf{74.3} & \textbf{78.5} & \textbf{58.6} & \textbf{72.9} & \textbf{78.8}\\
    & \ding{51} & 78.5 & 71.8 & 76.4 & 56.2 & 71.2 & 77.3\\
    \ding{51} &  & 80.7 & 73.6 & 77.9 & 58.1 & 72.3 & 78.4\\
    
    \midrule

    \hline
    \rowcolor[gray]{0.85}
    \multicolumn{2}{c|}{\modelname{}} & \multicolumn{6}{c}{Molecule Property Prediction} \\
    SGR & SEG & Tox21 & ToxCast & \multicolumn{1}{c}{Sider} & \multicolumn{1}{c}{ClinTox} & BBBP & Bace \\
    \midrule
    \ding{51} & \ding{51} & \textbf{77.5} & \textbf{61.4} & \multicolumn{1}{c}{\textbf{67.3}} & \multicolumn{1}{c}{\textbf{94.6}} & \textbf{72.6} & \textbf{84.2}\\
    & \ding{51} & 75.4 & 60.2 & \multicolumn{1}{c}{66.4} & \multicolumn{1}{c}{92.1} & 71.6 & 82.7\\
    \ding{51} &  & 76.2 & 60.8 & \multicolumn{1}{c}{66.9} & \multicolumn{1}{c}{94.0} & 72.1 & 83.4\\
    
    \bottomrule
    \end{tabular}
    \label{table:ablations}
\end{table}

\begin{figure}[b!]
    \centering
    \subfloat{\includegraphics[width=\linewidth]{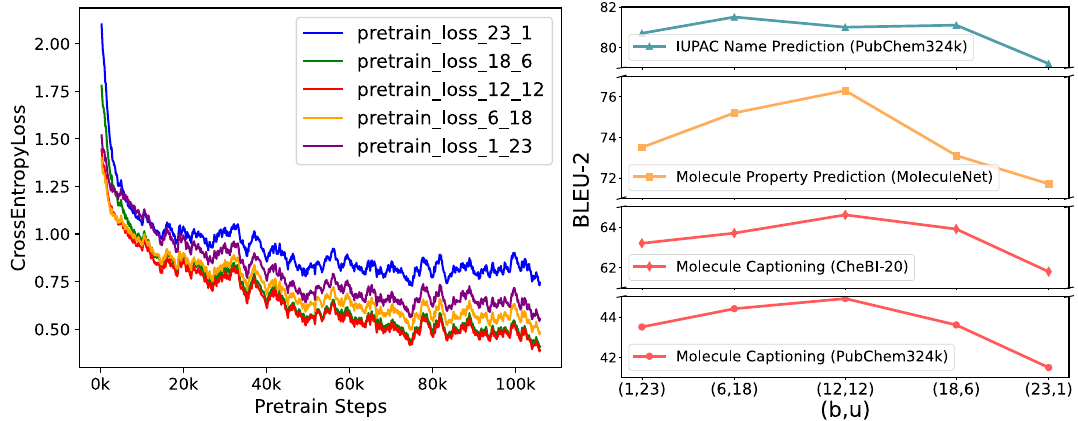}}
    \caption{The experimental results of different partition settings for the LLM. (Left) The pre-training loss curves. (Right) The test results of four molecular tasks.}
    \label{fig:low_up_layers}
\end{figure}

\paragraph{\textbf{Effectiveness of SEG}}
SEG enables converting any $b$ groups of structural embeddings $Z \in \mathbb{R}^{b \times w \times d}$ into specified $u$ groups of cross-view prefixes $\hat{Z} \in \mathbb{R}^{u \times w \times d}$, thus allowing the LLM to be partitioned arbitrarily.
However, when $b=u$, an alternative is to directly prepend $Z$ into the LLM's upper segment, namely discarding the SEG.
In this condition, we compare the performance of \modelname{} and $\text{\modelname}_{\text{w/o \textsc{seg}\xspace}}$, where $Z$ passes through and bypasses SEG, respectively. As shown in Table~\ref{table:ablations}, \modelname{} outperforms $\text{\modelname}_{\text{w/o \textsc{seg}\xspace}}$ across diverse tasks. 
Due to the varying structural features conveyed by each group of $Z$, SEG combines these groups to generate cross-view prefixes $\hat{Z}$, enabling each prefix to encapsulate comprehensive structural information and thereby enhancing the LLM's ability to understand molecular structures more accurately.

\paragraph{\textbf{Impact of Different LLM Partitions}}
Galactica consists of 24 layers in total. We compare five \modelname{} variants with $(b, u)$ set to $(1, 23)$, $(6, 18)$, $(12, 12)$, $(18, 6)$, and $(23, 1)$, respectively.
As illustrated in Figure~\ref{fig:low_up_layers}, $\text{\modelname}_{b=12,u=12}$ performs best on the PubChem324k and CheBI-20 molecule captioning datasets. 
Additionally, $\text{\modelname}_{b=6,u=18}$ performs best on the PubChem324k IUPAC name prediction dataset. 
Increasing $b$ provides more groups of SMILES guidance for SGR, facilitating the resampling process. Conversely, increasing $u$ allows more LLM layers to leverage cross-view prefixes, enhancing the utilization of structural information. There is a trade-off between $b$ and $u$ for different tasks, with generally better performance achieved when $b$ and $u$ are nearly balanced.

\begin{table}[t!]
    \centering
    \setlength{\tabcolsep}{3.8pt}
    \caption{The comparison in terms of computational cost across four models on the PubChem324k molecule captioning dataset. The metrics include the average length of cross-view prefixes, average FLOPs for these prefixes and training time on the PubChem324k molecule captioning dataset. 
    }
    \begin{tabular}{l|cccc}
    \toprule
    \hline
    \rowcolor[gray]{0.85}
    Model & \makecell{BLEU-2} & \makecell{Avg. Prefixes \\ (Count)} & \makecell{Avg. FLOPs \\ (Billions)} & \makecell{Train. Time \\ (Hours)} \\
    \hline
    \addlinespace[3pt]
    Galactica & 36.6 & - & - & 5.37 \\
    \addlinespace[-6pt]
    \multicolumn{5}{c}{\makebox[3cm]{\hdashrule{0.95\linewidth}{0.4pt}{1pt 1pt}}} \\
    \addlinespace[-1pt]
    $\text{\modelname}_{\textit{arch1}}$ & 42.8 & 288 & 353.52 & 7.14 \\
    $\text{\modelname}_{\textit{arch2}}$ & 41.5 & 20 & 24.20 & 5.86 \\
    $\text{\modelname}_{\textit{arch3}}$ & \textbf{44.9} & \textbf{10} & \textbf{14.05} & \textbf{5.53} \\
    \bottomrule
    \end{tabular}
    \label{table:speed}
\end{table}

\paragraph{\textbf{Efficiency Analysis}}
We compare the performance of \modelname{} with three different architectures, as shown in Figure~\ref{fig:arch_compare} (\textit{arch1}, \textit{arch2} and \textit{arch3}), under the setting of inputting $\boldsymbol{S}$, $\boldsymbol{G}$ and $\boldsymbol{I}$ simultaneously.
Galactica serves as the common baseline, processing only $\boldsymbol{S}$.
The results are shown in Table~\ref{table:speed}.
Compared to Galactica, three variants of \modelname{} achieve significant improvements.
Among them, $\text{\modelname}_{\textit{arch3}}$ achieves superior performance with shorter cross-view prefixes, fewer additional FLOPs, and smaller training time overhead.
Specifically, compared to $\text{\modelname}_{\textit{arch1}}$, $\text{\modelname}_{\textit{arch3}}$ reduces the length of cross-view prefixes by 96.5\%, the additional average number of FLOPs by 96.0\% and the training time by 22.5\%.
This demonstrates the architectural advantages of $\text{\modelname}_{\textit{arch3}}$, including utilizing SMILES guidance during the resampling to boost the quality of derived cross-view prefixes, and prepending prefixes to multiple LLM layers to promote structural information utilization.

%
%
\section{Conclusion}
In this work, we identify the fundamental limitations of relying solely on the molecular graph view, and propose \modelname{}, an innovative and scalable MLLM architecture that can integrate both topological and spatial structural views to jointly advance molecular understanding while maintaining computational efficiency. 
We primarily explore enhancing LLMs by integrating molecular graphs and images, which are representative topological and spatial views respectively, and highlight the impressive effectiveness of molecular images for enhancing the performance of LLMs in generative tasks.
In future research, we will consider fine-tuning \modelname{} on large-scale molecular instruction datasets, and integrating more representative molecular views into the \modelname{}.

\begin{acks}
This research was supported by the Strategic Priority Research Program of Chinese Academy of Sciences (XDA0490000) and the National Natural Science Foundation of China (62276245).
\end{acks}

\bibliographystyle{ACM-Reference-Format}
\balance
\bibliography{main}

\setcounter{secnumdepth}{2}
\cleardoublepage
\appendix
\renewcommand{\thesubsection}{\Alph{subsection}}

\twocolumn[
  \begin{center}
    \section*{\quad\quad\quad\quad\quad\quad\quad\quad\quad\quad\quad\quad\quad\quad\quad\quad\quad\quad\quad\quad\quad  Appendix}
  \end{center}
]
\subsection{Concatenate or Resample Molecular Modalities}
\label{app:expA}
\begin{figure}[h!]
    \centering
    \subfloat{\includegraphics[width=0.95\linewidth]{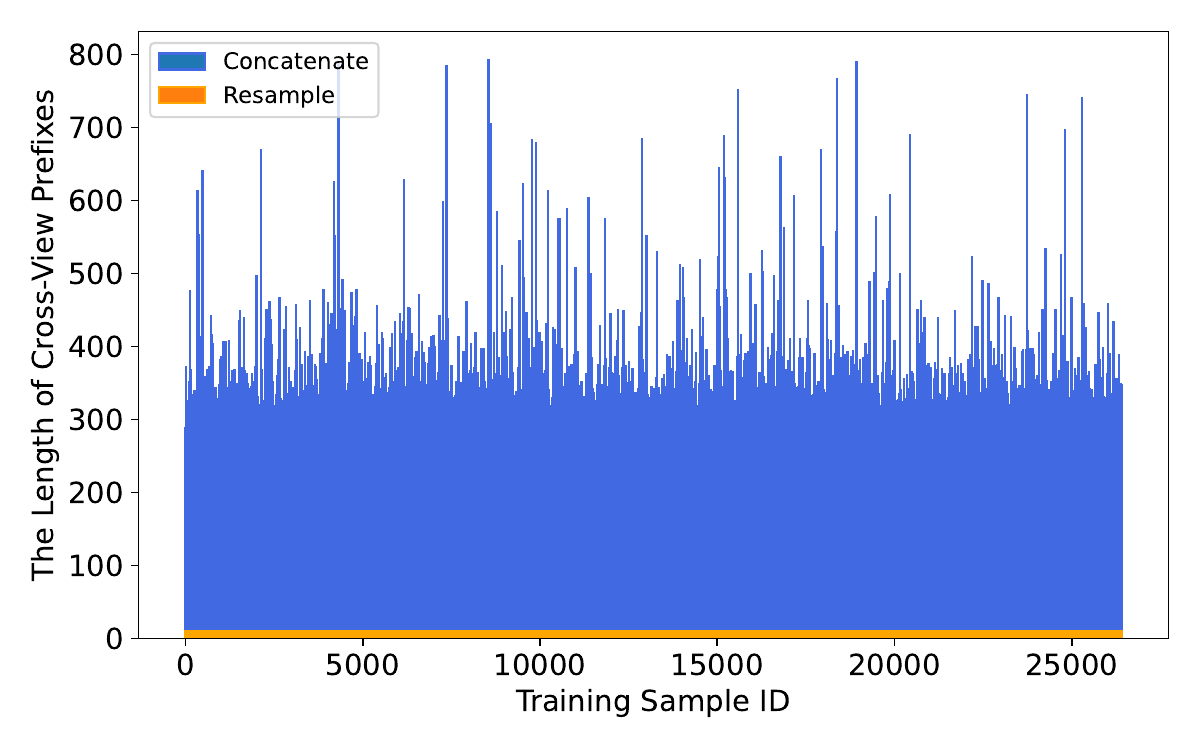}}
    \caption{The respective lengths of directly concatenated and resampled molecular graph and image embeddings on CheBI-20 molecule captioning training set.}
    \label{fig:graph_image_input}
\end{figure}

\begin{figure}[h!]
    \centering
    \subfloat{\includegraphics[width=0.7\linewidth]{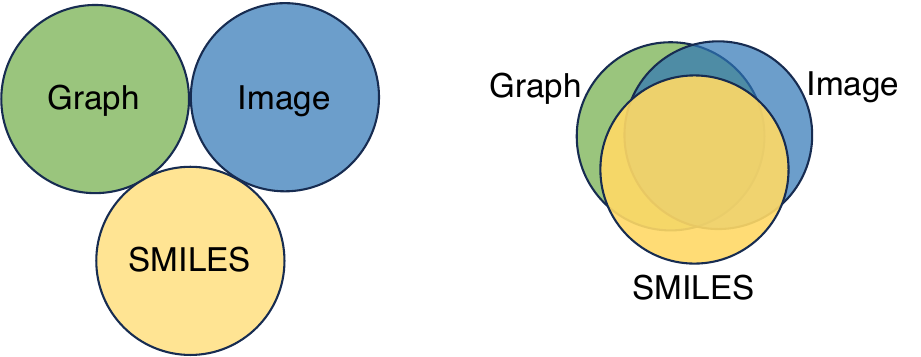}}
    \caption{(Left) Directly concatenate molecular SMILES, graph and image views. (Right) Take the information redundancy across views into account, and resample additional unique features from graph and image to augment SMILES representation.}
    \label{fig:three_mode_combine}
\end{figure}

In this research, we utilize additional graph and image views to enhance the molecular understanding of LLMs. Compared to SMILES, graph and image explicitly provide topological and spatial structural information of molecules, respectively. As illustrated in Figure~\ref{fig:graph_image_input}, if we employ the LLaVA architecture~\citep{liu2023visual}, directly concatenating graph and image embedding, and then inputting these embeddings along with SMILES, it will cause obviously increased input length. In fact, each form can basically represent a molecule alone. On this basis, each form possesses some unique strengths. Therefore, we expect to avoid information redundancy across the three forms and just combine the unique strengths of each to compress the input length, as illustrated in Figure~\ref{fig:three_mode_combine}. Considering that LLMs are inherently dependent on SMILES, which are essentially text-like sequences, we choose to augment this representation with resampled unique structural features from graph and image. Finally, with SMILES as guidance, molecular graph and image are resampled into a compact fixed-length embedding with unique information, and input along with SMILES to serve as a augmented molecular representation. This method reduces the additional input length by 96.5\%.


\subsection{Molecular Modalities}
\label{app:expB}
This research utilizes molecular SMILES, graphs and images. Molecular graphs and images could be obtained using the RDKit toolkit~\citep{bento2020open} according to SMILES. The following provides an overview of diverse molecular
modalities.

\noindent\textbf{SMILES.}
The Simplified Molecular Input Line Entry System (SMILES) is a specification that represents molecules using ASCII strings. Atoms are represented by letters, such as C for carbon, N for nitrogen and O for oxygen. Bonds are represented by specific characters, such as single (-), double (=), triple (\#), and aromatic (:) connections. Ring structures are denoted by assigning numbers to the involved atoms. In this research, we follow Galactica's approach and tokenize SMILES by character. For example, \( CCC(C)(C(=O)O)O \rightarrow C,C,C,(,C,),(,C,(,=,O,),O,),O. \) Currently, researchers also utilize SELFIES as molecular sequential representation. 

\begin{figure}[h!]
    \centering
    \subfloat{\includegraphics[width=\linewidth]{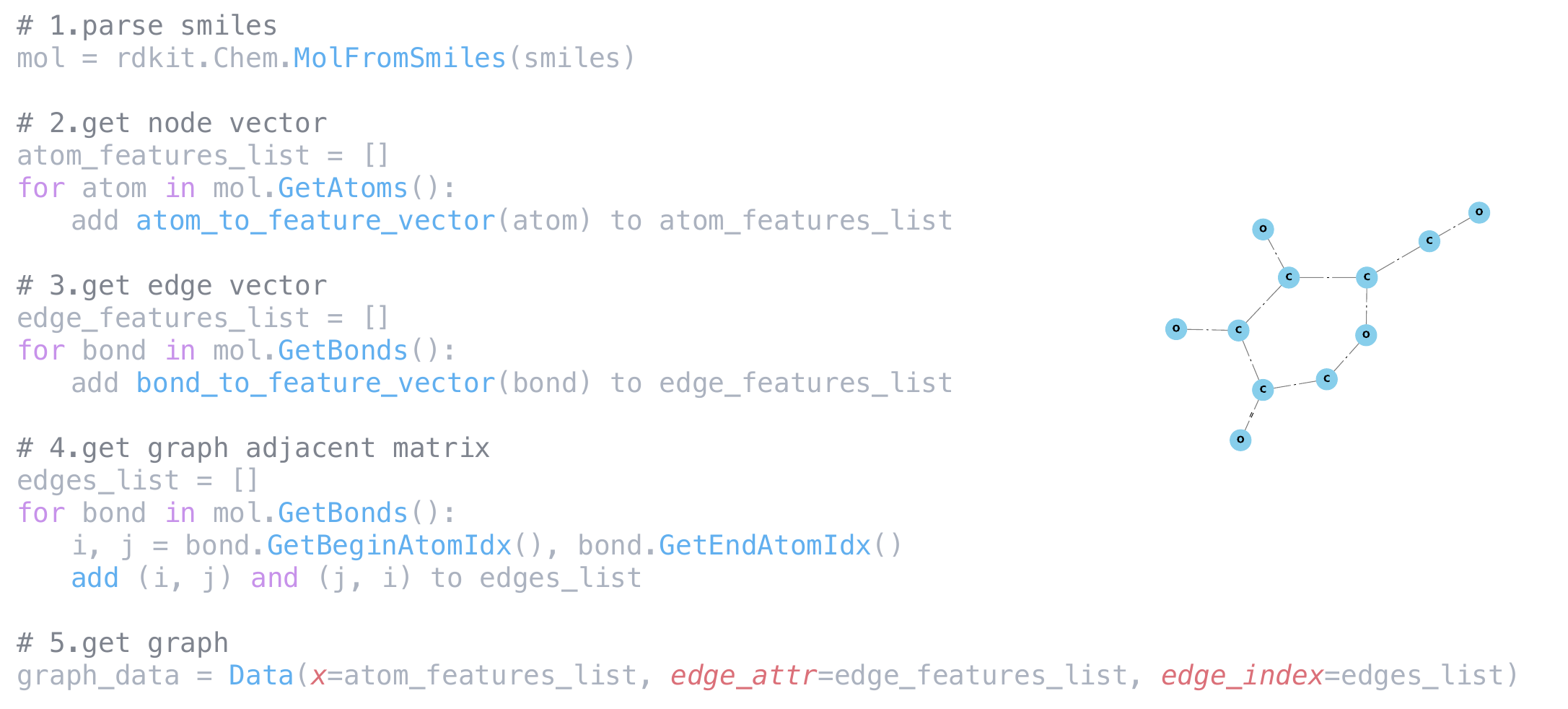}}
    \caption{Pseudo-code for building the molecular graph.}
    \label{fig:smiles_graph}
\end{figure}

\noindent\textbf{Graph.}
Graph represents the connectivity of atoms. In molecular graph, atoms serve as nodes and bonds serve as edges. Each node is attributed with atomic type, hybridization state, formal charge, etc. Each edge can be attributed with bond type, bond order, etc. In graph modeling, node attributes are assigned corresponding vectors from embedding matrix. Node connection relationships are denoted by adjacent matrix. Neighboring nodes exchange information through the message passing mechanism. We utilize the RDKit toolkit to build the molecular graph based on SMILES, and the pseudo-code is illustrated in Figure~\ref{fig:smiles_graph}.

\begin{figure}[h!]
    \centering
    \subfloat{\includegraphics[width=\linewidth]{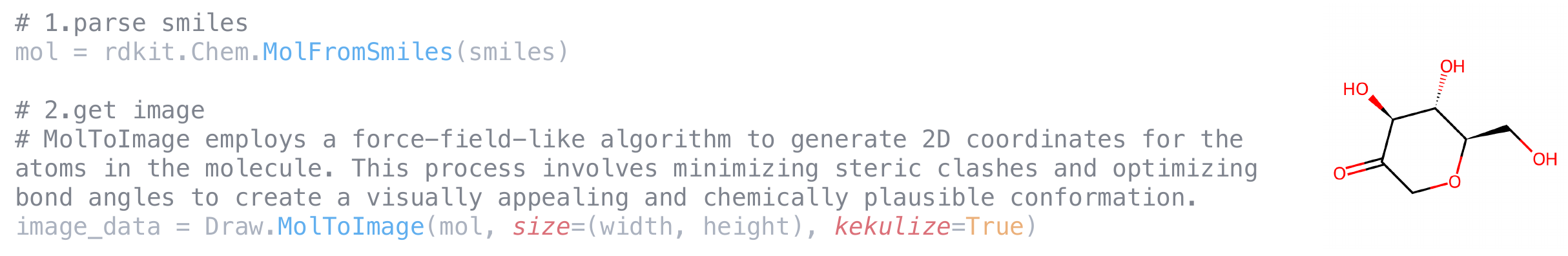}}
    \caption{Pseudo-code for building the molecular image.}
    \label{fig:smiles_image}
\end{figure}

\noindent\textbf{Image.}
Image represents the molecular overall shape in space. In molecular image, atoms and bonds are typically depicted as pixels. Atom and bond types can be denoted through variations in color or intensity, but fine-grained attributes such as conformational flexibility and electronic properties are difficult to represent visually. We utilize the RDKit toolkit to build the molecular image based on SMILES, and the pseudo-code is illustrated in Figure~\ref{fig:smiles_image}. We ensure the consistency of generated molecular images by keeping the input parameters of the \textit{MolToImage()} function the same, such as \textit{kekulize=True} and \textit{wedgeBonds=True}.

\subsection{Impact of SMILES Guidance Length}
\label{app:expC}
\begin{figure}[h!]
    \centering
    \subfloat{\includegraphics[width=0.9\linewidth]{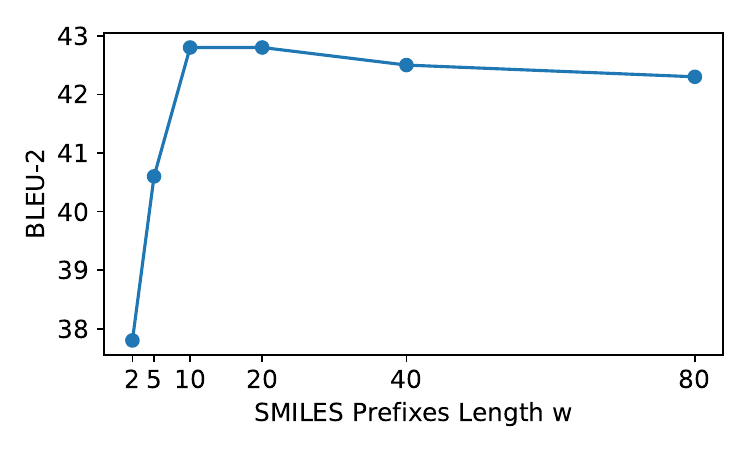}}
    \caption{The performance of \modelname{} with different prefixes length $w$ on PubChem324k molecule captioning dataset.}
    \label{fig:prefixes_length}
\end{figure}

In this section, we investigate the impact of SMILES guidance length on \modelname{}'s performance and determine the final SMILES guidance length $w$ used in the main experiments. We conduct the analysis on PubChem324k molecule captioning dataset and adopt the setting of utilizing molecular SMILES, graphs and images simultaneously.  
The experiment results are illustrated in Figure~\ref{fig:prefixes_length}.
When the prefixes length $w$ increases from 2 to 10, there is a notable improvement. This is due to the fact that excessively short SMILES guidance severely restrict the resampling of molecular graph and image modalities, leading to substantial information loss. Appropriately increasing $w$ will greatly improve the performance.
When the length of prefixes reaches 10, further increasing $w$ has a marginal impact on \modelname{}'s performance, and may even slightly damage it. This is attributed to the escalating number of learnable prefixes potentially causing the model to overfit to the training set. Based on this analysis, we finally set $w$ as 10 in our main experiments.

\noindent\textbf{Other Molecular Modalities.}
In 3D molecular graph \citep{halgren1996merck,li2021hamnet,lu2019molecular,klicpera2021gemnet,fang2022geometry,liu2021pre}, atomic coordinates are considered, allowing for accurate representations of molecular geometry. In 3D molecular grid\citep{xie2020multitask,sunseri2020libmolgrid,liu2019multiresolution,casey2020prediction,tran2019deepnose,kuzminykh20183d}, the space surrounding the molecule is divided into a regular grid, and each grid point is assigned a value reflecting molecular properties.

\subsection{Experimental Details}
\label{app:expD}
\noindent\textbf{Large Language Models.}
In this research, we adopt Galactica~\citep{taylor2022galactica}, a specialized large language model developed by Meta AI for the science domain, as \modelname{}'s language backbone. Other LLMs with decoder-only architecture, such as Llama~\citep{touvron2023llama}, can also be effortlessly adopted in \modelname{}. Galactica is pretrained on a comprehensive and high-quality scientific corpus that includes over 48 million papers, textbooks, as well as data from PubChem Compound, UniProt, and scientific websites. Galactica includes several variants with size ranging from 125 million to 120 billion. Due to the limitation of computational resources, we adopt $\text{Galactica}_\text{1.3B}$ in this research.

\noindent\textbf{Pretrain Settings.}
In the pretraining stage, \modelname{} performs the captioning task based on molecular SMILES, graphs and images. Trainable modules in \modelname{} (including five variants with different partitions) include the SMILES guidance, SMILES Guided Resampler and Structural Embedding Gate. Trainable modules in $\text{\modelname}_{\text{w/o \textsc{sgr}\xspace}}$ include the random embeddings, normal resampler, and Structured Embedding Gate. Trainable modules in $\text{\modelname}_{\text{w/o \textsc{seg}\xspace}}$ include the SMILES guidance, and SMILES Guided Resampler. The length of prefixes $w$ in the LLM is set as 10. The number of layers in SMILES Guided Resampler is set as 4. The total number of pretraining epochs is set as 20. We adopt the AdamW optimizer and set the weight-decay as 0.01. The cosine learning rate scheduler is adopted with a warmup of 1000 steps and a peak learning rate of 2e-4. 

\noindent\textbf{Finetune Settings.}
On molecule captioning and IUPAC name prediction task, in addition to the trainable modules mentioned in the pretraining stage, we unfreeze the molecular encoders and utilize LoRA to finetune the LLM. We set LoRA rank $r=32$, $\alpha=64$, and the target modules of LoRA include the LLM's \textit{q\_proj}, \textit{v\_proj}, \textit{out\_proj}, \textit{fc1} and \textit{fc2}. 
On molecule property prediction task, LoRA is not utilized. 
Other hyperparameters are set as in the pretraining stage.
In all experiments, we finetune \modelname{} for 100 epochs and select the model with highest BLEU-2 score or ROC-AUC score on validation set for testing.

\subsection{Prompts}
\begin{figure}[h!]
    \centering
    \subfloat{\includegraphics[width=\linewidth]{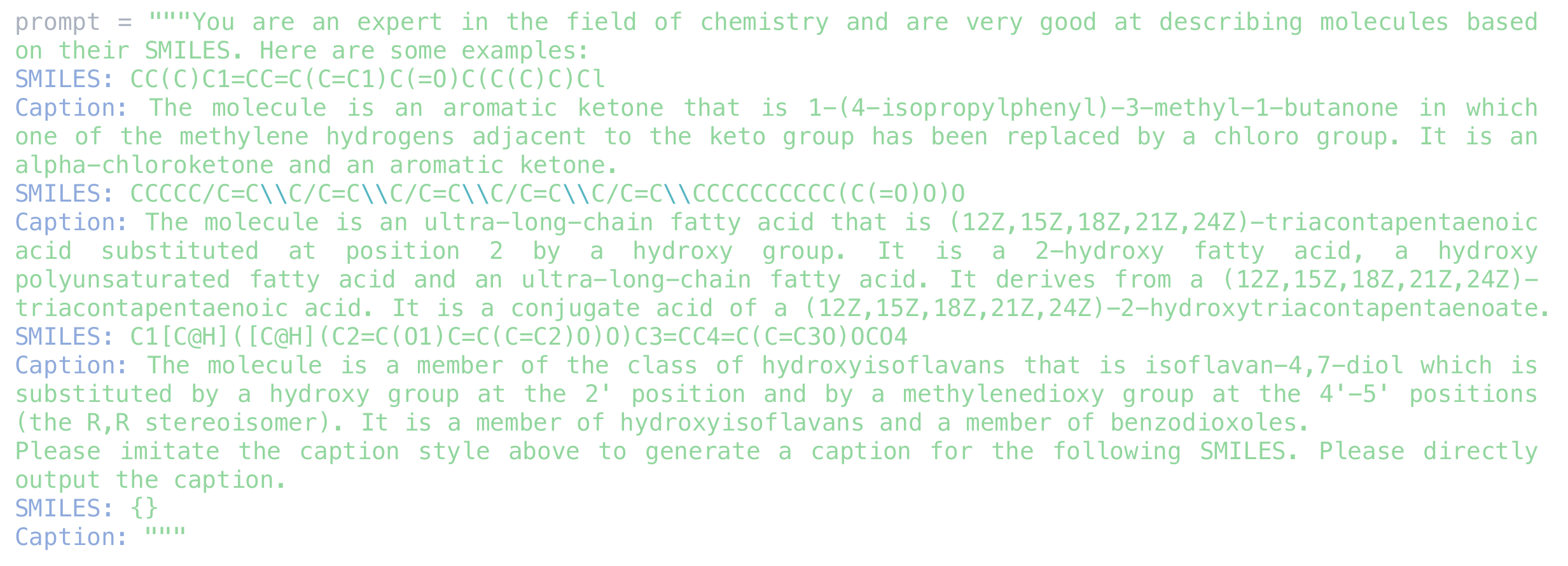}}
    \caption{Prompt for evaluating the performance of GPT-4o and $\text{Llama3}_{\text{8B-Instruct}}$ on the molecule captioning task.}
    \label{fig:promt_caption}
\end{figure}

\begin{figure}[ht!]
    \centering
    \subfloat{\includegraphics[width=\linewidth]{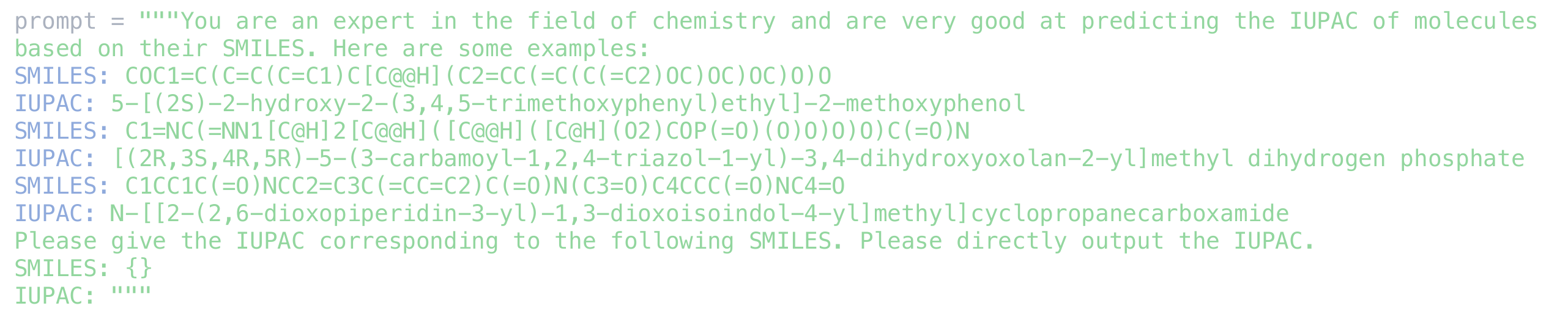}}
    \caption{Prompt for evaluating the performance of GPT-4o and $\text{Llama3}_{\text{8B-Instruct}}$ on the IUPAC name prediction task.}
    \label{fig:prompt_iupac}
\end{figure}

We provide the prompts utilized in the experiments, as illustrated in Figure~\ref{fig:promt_caption} and Figure~\ref{fig:prompt_iupac}

\end{document}